\documentclass[twocolumn,preprintnumbers,amsmath,amssymb]{revtex4}

\usepackage{amsmath}
\usepackage{amssymb}
\usepackage{epsfig}
\begin{document}

\title{Active sub-Rayleigh alignment of parallel or antiparallel laser beams}

\author{Holger M\"uller, Sheng-wey Chiow, Quan Long, Christoph Vo, and Steven Chu.\\
Physics Department, Stanford University, Stanford, CA 94305}

\begin{abstract}
We measure and stabilize the relative angle of (anti-)parallel laser beams to 5\,nrad/$\sqrt{\rm Hz}$ resolution by comparing the phases of radio frequency beat notes on a quadrant photodetector. The absolute accuracy is 5.1\,$\mu$rad and 2.1\,$\mu$rad for antiparallel and parallel beams, respectively, which is more than 6 and 16 times below the Rayleigh criterion. 
\end{abstract}
\maketitle

Lasers are universal tools as high-precision ``rulers", i.e., references of length, time, and/or spatial direction. Examples are the definition of the meter, experiments in fundamental physics \cite{KT,MM}, and precision inertial sensing. Extremely high accuracy, both in terms of frequency and phase \cite{MuellerPLL} as well as pointing stability, is demanded in atom interferometry \cite{Chu,ChuHouches,Berman}, which is applied for atomic clocks and inertial sensing for the measurement of Newton's constant $G$, the local gravitational acceleration $g$, and the ratio of the Planck constant to the mass of an atom $\hbar/M$. In atom interferometers that use two-photon transitions driven by counterpropagating laser pulses as beam splitters \cite{Kas91a,Kas91b}, the phase of the matter waves is measured against the effective wavevector $|\mathbf k_{\rm eff}|=|\mathbf k_1|-|\mathbf k_2|\cos \alpha$, where $\alpha\simeq \pi$ is the angle between the individual wavevectors $\mathbf k_{1,2}$. Present precision atom interferometers, such as the measurement of $\hbar/M$ to a few $10^{-9}$ absolute accuracy, require the same level of accuracy in $|\mathbf k_{\rm eff}|$ \cite{Wicht}, and further improvements are expected. For an accuracy goal of $10^{-10}$, initial misalignment, vibration, or creep in the counterpropagation angle $\alpha$ must be kept below $15\,\mu$rad. However, creep of conventional optics setups leads to a systematic error of the order of $10^{-9}$ in $|\mathbf k_{\rm eff}|$ in present experiments \cite{Wicht}. Therefore, we developed an active control system to stabilize the alignment of (anti-)parallel laser beams. 

The power of optical systems to resolve a small angle $\beta$ between light rays is often given in terms of the Rayleigh criterion $\beta\gtrsim \beta_R\equiv \lambda/a$, where $\lambda$ is the wavelength and $a$ the aperture \cite{Rayleigh}. Even for the large beam diameters common in atom interferometry (in our setup, $\lambda=0.852\,\mu$m and the beam waist $w_0\approx a/\pi \approx 0.8$\,cm, which implies $\beta_R\approx 34\mu$rad), sub-Rayleigh alignment is necessary for a $10^{-10}$ precision in $|\mathbf k_{\rm eff}|$. Conventional methods of measuring the angle $\beta$ are not well suited for this purpose: For example, a pair of pinholes is often used to test the alignment, but it does not indicate the direction of a misalignment. 
Active alignment is facilitated by using a corner cube to retroreflect one of the beams, and testing the parallelity afterwards. (Hollow corner cubes that are certified to have sub-arc second ($\approx 5$\,$\mu$rad) alignment errors are commercially available.) High sensitivity is achieved by using the interference fringes between the beams to indicate the relative alignment \cite{Hobbs,Morrison,Heinzel}. 

\begin{figure}
\centering
\epsfig{file=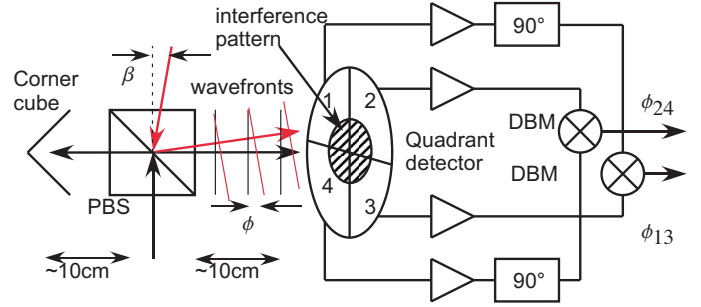, width=0.5\textwidth}
\caption{\label{BPMschema} Schematic of the setup for antiparallel beams.}
\end{figure}

In this letter, we report on a system that solves the problem of beam pointing fluctuations in optics setups such as the laser systems used in atom interferometry. We detect the relative angle by a beat note measurement on a quadrant photodetector (QD) (Fig. \ref{BPMschema}), using radio frequencies above the technical $1/f$ noise floor of lasers. 
A servo tracks the pointing of one beam relative to the other. The high absolute accuracy of our method is confirmed by a detailed study of the systematic effects.\\

Whenever two overlapped electromagnetic waves having a frequency difference of $\omega$ are detected (using a photodiode, for example), interference causes an oscillating component (``beat note") at $\omega$. For an infinitely small detector area, the beat note's phase is equal to the phase difference $\phi$ of the waves at the location of the detector. For parallel beams, $\phi$ is constant on the cross section of the beams. If the beams are misaligned by an angle $\beta\ll 1$, however, $\phi= 2\pi r \beta/\lambda$ is a function of the distance $r$ from the center on the plane of incidence (Fig. \ref{BPMschema}). Thus, measuring $\phi(r)$ at different locations reveals the angular alignment. For separately measuring the two relevant angles, we measure the phases between two pairs of detectors. Therefore, we split off a $\sim 1\%$ intensity sample using a residual reflection from a polarizing beam splitter (PBS), where the polarizations are set for maximum transmission (Fig. \ref{BPMschema}). 
The sample beams are directed to a QD, one by retroreflection from the corner cube.

\begin{figure}
\centering
\epsfig{file=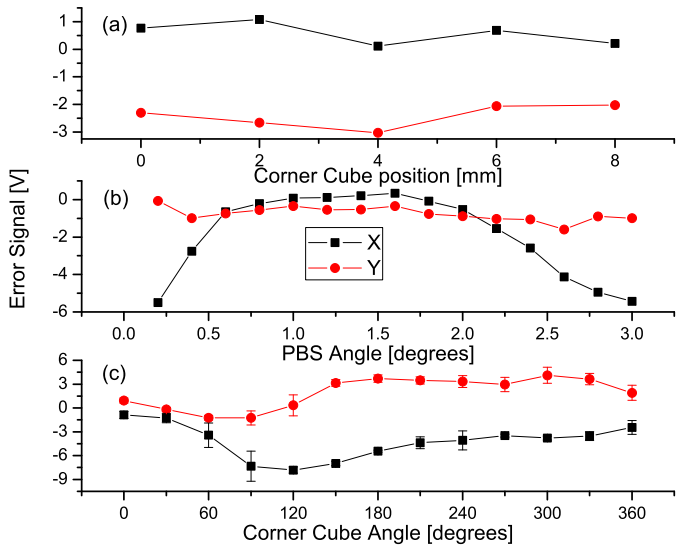,width=0.5\textwidth}
\caption{\label{tests} Tests for systematic influences. (a), parallel shift of the corner cube; (b), PBS alignment; (c), rotation of the corner cube. 
}
\end{figure}

For calculating the phase differences for a QD of finite radius $R$, we calculate the interference pattern due to two Gaussian beams within the Rayleigh range (where the wavefronts are essentially flat), tilted relative to each other by an angle $\beta\ll 1$. The beat note is given by the intensity integrated over the area of one quadrant
\begin{equation}
I_1=2|E_1E_2|\int_0^R\int_{-\pi/4}^{\pi/4}re^{-2r^2/w_0^2}\cos(\beta kr\cos\theta-\omega t)d\theta dr\,,
\end{equation}
where $r$ and $\theta$ are cylindrical coordinates on the QD's surface and $k=2\pi/\lambda$. The integration is carried out to the first order in $\beta$ by decomposing the outer cosine according to $\cos\alpha=(e^{i\alpha}+e^{-i\alpha})/2$ and then using $\exp(iz\cos\theta)=\sum_{n=-\infty}^\infty i^n J_n(z)e^{i n\theta}$. We write the result as $I_1={\rm const}+N\cos(\omega t+\phi_1)$, where $N$ is the amplitude of the beat note and $\tan \phi_1=(w_0/\lambda) s(\eta)\beta$ its phase. The function $
s(\eta)=2(\sqrt{\pi}\Phi(\eta)-2\eta e^{-\eta^2})/(1-e^{-\eta^2})$ depends on the ratio $\eta=\sqrt{2} R/w_0$ [$\Phi(\eta)=(2/\sqrt{\pi})\int_0^\eta e^{-t^2}dt$ is the probability integral]. For $w_0=0.8$\,cm, $R=0.5$\,cm, the phase difference between opposite quadrants $\phi_{13}=\phi_1-\phi_3=2\phi_1\sim 6.6\times 10^4 \beta$.

Since the beat is completely determined by the interfering radiation, its phases are quite insensitive to the arrangement of the detection system. Thus, $\phi_{13}=\phi_{24}=0$ indicates counterpropagation, regardless of the orientation of, e.g., the detector, the corner cube, or the PBS with respect to the beams (assuming no corner cube errors and that the setup is small compared to the wavelength of the beat frequency). Also, if the interfering beams are parallel, but not accurately overlapped, the phase shift between the quadrants remains zero, even if additionally the center of the QD is off the center of the interference pattern. Also note that $\phi_{13}$ and $\phi_{24}$ are independent of the rf frequency and that imbalances in photodetector sensititivity or area should not offset the zero of the phase measurement. This insenitivity to systematic effects and the linear dependence of the signal (rf phase difference) on the counterpropagation angle $\beta$ make our method well suited for reaching sub-Rayleigh absolute accuracy. 

The method directly stabilizes the relative orientation of the wavefronts, which is the quantity that matters in atom interferometry. As long as the detector is close to the atom interferometer, different radii or positions of the waists can be tolerated: If for each beam both the QD and the atoms are within $z_R/n$ distance to the waist and centered to within $w_0/m$, then the error will be $\lesssim\beta_R/(nm)$. For the above beam parameters and $n,m$ at least 5 (which is easy to achieve), $\beta_R/(nm)=1.5\,\mu$rad.

\begin{figure}
\centering
\epsfig{file=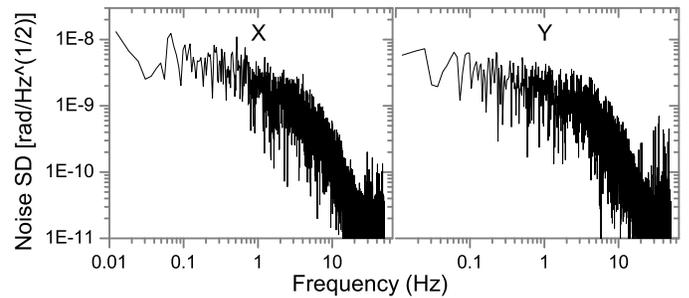,width=0.5\textwidth}
\caption{\label{noise} Noise spectral density (SD) of the angle measurement.}
\end{figure}

The optical setup is built on a (floating) optical table, using standard lens holders and mirror mounts. The signals from the quadrants are amplified by trans-impedance amplifiers in the cascode transistor configuration \cite{Hobbs}, using LM7171 operational amplifiers. The amplifier bandwidth of 40\,MHz reduces their phase shifts to 40\,mrad at $\omega=2\pi\times1\,$MHz, corresponding to a 0.4\,$\mu$rad offset in $\beta$; this, however, cancels out if the amplifiers are alike. The amplified signals are converted to ECL signals by four comparators (type AD96687). This reduces the influence of laser power variations on the subsequent stages. Double-balanced mixers (DBMs), type LPD-1 by Mini-Circuits, are used as phase detectors. Since DBMs produce zero average output voltage when driven by quadrature signals, we shift $\phi_1$ and $\phi_4$ by 90$^\circ$ using critically damped second-order low-pass filters before the comparators. Using broadband $90^\circ$ shifting networks such as quadrature power splitters, a frequency range of about 2:1  can be obtained (also removing the need for trimming). Using sample-and-hold type phase detectors would eliminate the need for a phase shifter and could have several GHz bandwidth. The dc output of the mixers is low-pass filtered to $\sim10$\,kHz bandwidth and amplified by a factor of 10 to $\pm 12$\,V range. The sensitivity of this circuit to the rf phase difference is measured to be 10.9\,mV/mrad for the X channel and 10.2\,mV/mrad for Y, corresponding to a sensitivity to the beam misalignment angle $\beta$ of 0.72\,V/$\mu$rad and 0.67\,V/$\mu$rad, respectively.

For testing and trimming the electronics, one could rotate the QD around its axis, which should cause no error in the counterpropagation angle. A simpler method, however, is to use two laser beams separated by 1\,MHz in frequency and overlapped in a common single-mode fiber as a copropagation reference. After an initial trim of the phase shifters, the drift is below 0.15\,$\mu$rad beam alignment error over several weeks. Reducing the intensity of the beams by a factor of up to 8 causes an error below $0.7\,\mu$rad. (Further reduction causes  errors as the beat notes become too weak for accurate phase measurement.) Thus, the overall long-term error of the electronics is estimated as $<1\,\mu$rad. 

\begin{figure}
\centering
\epsfig{file=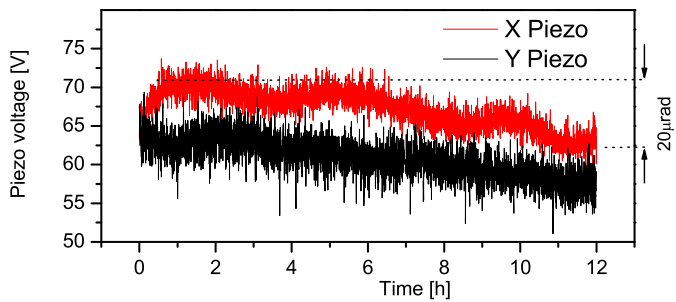,width=0.5\textwidth}
\caption{\label{timetrace} 12\,h operation of the system.}
\end{figure}

We also study the influence of the optical elements on the accuracy. Two counterpropagating beams are aligned to produce zero misalignment error signals. Setting an iris in front of the QD then causes no offset larger than 0.5\,$\mu$rad for diameters of 5-10\,mm. (For smaller diameters, errors arise due to insufficient signal amplitude.) Shifting the corner cube orthogonally to the sample beam directions by up to 8\,mm causes an error below 1.4$\,\mu$rad, see Fig. \ref{tests} (a). We test rotating the PBS away from its optimum position. For rotations below $\sim 0.8^\circ$, the error is below 1.3\,$\mu$rad, see Fig. \ref{tests} (b). Larger rotations cause strongly reduced signal amplitude: A 1$^\circ$ rotation offsets the centers of the beams on the QD by 3.3\,mm from each other and by 5\,mm - the QD's radius - from the QD (Fig. \ref{BPMschema}). Thus, this measurement also demonstrates the low influence of beam displacement. The error $\epsilon$ in the retroreflection angle of the corner cube is tested by rotating the corner cube around the beam axis, which should cause a modulation of $2\epsilon$ peak to peak in the X and Y alignment error signals. The data shown in Fig. \ref{tests} (c) shows a peak deviation of $\epsilon=4.6\,\mu$rad (including $\sim 2\,\mu$rad drift of the beam alignment during the measurement), within the manufacturer's specification of $\epsilon \leq 5\,\mu$rad. We also test the resolution by using two beams from a single mode fiber as a stable copropagation reference. The noise spectral density of both channels (Fig. \ref{noise}) is essentially white at $\lesssim 5$\,nrad/$\sqrt{\rm Hz}$ between 0.01 and 1\,Hz. 

For active control of the alignment by a proportional-integral (PI) feedback, we use a pair of mirrors that can be tilted by about 200\,$\mu$rad using piezo actuators with a sensitivity of $\sim 2\,\mu$rad/V and a resonance frequency of 1.2\,kHz. This works reliably, as demonstrated by the 12\,h time-trace shown in Fig. \ref{timetrace}. 
The servo corrects for short and long term alignment fluctuations of the order of 20\,$\mu$rad with $\sim$3\,nrad/$\sqrt{\rm Hz}$ residual noise in the error signals. In actual atom interferometry applications, parts of the optics may be mounted on a floating vibration isolator, whose creep would considerably increase the long term errors without active stabilization. We plan to operate the beam stabilization for about one second every few minutes and store the applied correction. This should be sufficient for keeping the error below 15\,$\mu$rad at all times. 

We have demonstrated a system for measuring and maintaining the counterpropagation of laser beams, based on retroreflecting one beam and comparing the phase of beat notes between them on a quadrant photodetector. We reach a resolution of 5\,nrad in 1\,s integration time. Tests for systematic influences 
indicate an overall absolute accuracy of better than 5.1\,$\mu$rad, six times below the Rayleigh criterion of 34\,$\mu$rad. If the setup is used for copropagating beams, the corner cube inaccuracy is eliminated, giving 2.1\,$\mu$rad absolute accuracy. 

H. M\"uller's email address is holgerm@stanford.edu. This work is sponsored in part by grants from the AFOSR, the NSF, and the MURI. H.M. acknowledges the support of the Alexander von Humboldt Foundation. C.V. acknowledges the support by the DAAD and the Stiftung der Deutschen Wirtschaft.


\begin{thebibliography}{99}
\bibitem{KT} C. Braxmaier, H. M\"uller, O. Pradl, J. Mlynek, A. Peters, and S. Schiller, Phys. Rev. Lett {\bf 88,} 010401 (2002).

\bibitem{MM} H. M\"uller, S. Herrmann, C. Braxmaier, S. Schiller, and A. Peters, Phys. Rev. Lett. {\bf 91,} 020401 (2003).

\bibitem{MuellerPLL} H. M\"uller, S.-w. Chiow, Q. Long, and S. Chu, physics/0507187 (2005).

\bibitem{Chu} S. Chu, Nature (London) {\bf 416}, 206 (2002).

\bibitem{ChuHouches} S. Chu in: {\em Coherent atomic matter waves}, R. Kaiser, C. Westbrook, F. David (editors) (Springer-Verlag, Berlin, 2001), 317-370.

\bibitem{Berman} Paul Berman, {\em Atom Interferometry} (Academic Press, New York, 1997).

\bibitem{Kas91a} M. Kasevich, D. Weiss, E. Riis, K. Moler, S. Kasapi, and S. Chu, Phys. Rev. Lett. {\bf 66}, 2297 (1991).

\bibitem{Kas91b} M. Kasevich and S. Chu, Phys. Rev. Lett. {\bf 67,} 181 (1991).

\bibitem{Wicht} A. Wicht, J. M. Hensley, E. Sarajlic, and S. Chu, Physica Scripta {\bf T102}, 82 (2002).

\bibitem{Rayleigh} Lord Rayleigh, Phil. Mag. {\bf 8}, 261-274 (1879).


\bibitem{Hobbs} Philip C.D. Hobbs, {\em Building Electro-Optical Systems} (Wiley, New York, 2000).

\bibitem{Morrison} E. Morrison, B.J. Meers, D.I. Robertson, and H. Ward, Appl. Opt. {\bf 33}, 5041 (1994).

\bibitem{Heinzel} G. Heinzel, V. Wand, A. Garcya, O. Jennrich, C. Braxmaier, D. Robertson, K. Middleton, D. Hoyland, A. R\"udiger, R. Schilling, U. Johann, and K. Danzmann, Class. Quantum Grav. {\bf 21} S581-S587 (2004).

\end{thebibliography}
\end{document}